\documentclass[preprint]{aastex}
\usepackage{graphicx}
\usepackage{lscape}

\usepackage{times}
\usepackage{epsfig}
\usepackage{epsf}
\usepackage{rotating}

\begin{document}


\title{Detection of highly ionized CIV gas within the Local Cavity.}


\author{
Barry Y. Welsh\altaffilmark{1},
Jonathan Wheatley \altaffilmark{1}, Oswald H.W. Siegmund \altaffilmark{1}
and Rosine Lallement \altaffilmark{2}}

\altaffiltext{1}{Space Sciences Laboratory, University of California, 7 Gauss Way, Berkeley, CA 94720, USA} \altaffiltext{2}{IPSL / LATMOS, Versailles, France}



\begin{abstract}
We present high resolution (R = 114,000) ultraviolet measurements of the 
interstellar absorption line profiles
of the CIV (1550\AA) high ionization doublet  recorded towards the nearby B2Ve star HD 158427 (d $\sim$ 74pc).
These data, which were recorded with the recently re-furbished STIS instrument on the $\it HST$, represent
the most convincing detection yet of highly ionized CIV absorption that can be associated with interstellar gas located $\it within$
the boundary of the Local Cavity. Two highly ionized gas clouds at V$_{1}$ = -24.3 km s$^{-1}$ and V$_{2}$ = -41.3 km s$^{-1}$
are revealed in both CIV absorption lines, with the V$_{1}$ component almost certainly being due to absorption by
the Local Interstellar Cloud ( d $<$ 5pc). Although the observed column densities for both cloud components can be explained
by the predictions of current theoretical models of the local interstellar medium, the narrow doppler
width of the  V$_{2}$ line-profile ( $\it b$ = 6.8 km s$^{-1}$)
indicates an unusually low gas  temperature of $\le$ 34,000K) for this highly ionized component. It is conjectured that
the V$_{2}$ cloud may be due to an outflow of highly ionized and hot gas from the nearby Loop I superbubble. These new
data also indicate that absorption due to highly ionized gas in the Local Cavity can be best described as being `patchy' in nature.
\end{abstract}



\maketitle


\section{Introduction}
The volume of the interstellar medium that surrounds the Sun to a radius of $\sim$ 80pc in most directions,
often called the Local Cavity (LC),
is known to possess an unusually low neutral gas density of n$_{HI}$ $<$ 0.1 cm$^{-3}$. However, the physical
characteristics of the plasma that fills
this interstellar void are still a matter of vigorous debate. The 
traditional view
of this rarefied interstellar region is that it is filled with a soft X-ray emitting plasma with a temperature of
$\sim$ 1 million K \citep{snowden98}, with the additional presence of numerous interspersed warm (T $\sim$ 7000K)
and partially ionized diffuse cloudlets \citep{lall86,redf08}. However, over the past decade several
observational inconsistencies have questioned the validity of this picture of a local hot and highly ionized interstellar bubble.
For example, solar wind charge exchange reactions in the heliosphere have been shown to produce a significant
fraction of the soft X-ray emission that was previously ascribed to the hot Local Cavity \citep{robert03, kout09},
and extreme ultraviolet line emission predicted by several models of a hot Local Cavity has not been detected
by both the $\it EUVE$ and $\it CHIPS$ satellite missions \citep{jel95,hurw05}. For a fuller discussion of
the `hot or not' Local Cavity debate we refer readers to Welsh $\&$ Shelton \cite{ welsh09}.

Strong evidence for a local highly ionized plasma with a (lower) temperature of $\sim$ 3 x 10$^{5}$K
has come from observations of the OVI (1032\AA) ion, seen in the far UV absorption spectra
of several stars located within 100pc \citep{jenk78, sav06}. This ion is generally thought to trace collisionally
ionized gas that can exist in (turbulent) transition zones
between the hot million K plasma of the LC and cooler gas clouds residing therein.  In addition, one may
also expect to detect local absorption from the high ion line doublets of NV (1238\AA), CIV (1550\AA) and
SiIV (1394\AA),
which can form in lower temperature gas at $\sim$ 10$^{5}$K. Several theoretical models have been proposed
to explain the observed ratios of the lines of OVI, NV, CIV and SiIV for more distant regions of the
interstellar medium (ISM), that invoke shocks, photoionization,  evaporating interfaces or rapidly cooling
gas \citep{indeb04}. Typically the population ratios of NV and CIV to OVI lie in the range 0.1 - 0.4,
such that if OVI is well detected then the other two ions should also be present. However, there is currently only one sight-line within
100pc  (towards $\alpha$ Vir, d = 80pc) against which detections of
interstellar NV, CIV and SiIV absorption can be tested against theory \citep{huang95}. Although
these observed column density ratios are broadly consistent with the theory of Slavin $\&$ Cox \citep{slav92} in which cooling hot gas
condenses onto the interior shell of an evolving supernova remnant bubble, we note that
this (high latitude) star has a significant surrounding HII region, as traced by its H-$\alpha$ and SII
emission contours \citep{haff03}.  These high ion absorption lines are all centered near the rest frame
velocity of the star, suggesting a circumstellar rather than an interstellar origin.
Therefore, other production mechanisms such as  ionization
of the surrounding HII region by hot X-ray emitting gas
cannot be discounted for this sight-line \citep{cow81}. 

One universally accepted physical characteristic of interstellar gas within the LC is the ubiquitous presence of
photoionized clouds with a temperature of $\sim$ 10$^{4}$K \citep{red04, lehn03}.  The most well-observed of
these clouds is the Local Interstellar Cloud (LIC) that surrounds most of the heliosphere \citep{lall95}.  The
LIC is a partially ionized structure with a temperature of  $\sim$ 6500K, an electron density of 0.07 cm$^{-3}$ with
elements such as Fe and Mg being depleted with respect to solar values \citep{red09}. 
 Much effort has been directed into understanding the ionization state
and the potential sources and radiative transfer of ionization for the LIC \citep{bruh88, slav02}. The unusual ionization of
the local ISM (in which He is more ionized than H) suggests that the LIC may be out of ionization equilibrium \citep{lyu96, wolff99},
whereas local observations of ArI and OI indicate that the LIC is close to ionization
and thermal equilibrium \citep{jenk00}. 

A firm prediction of many of these ionization models of the local gas is that the CIV ion should be formed at
the evaporative interface between the LIC and the purported ambient
hot plasma of the LC \citep{bohr87, slav02}. Such models typically predict CIV column density
values of N(CIV) = 1 - 3 x 10$^{12}$ cm$^{-3}$ per interface, which translates into absorption
equivalent widths which should be measurable by present UV instrumentation on telescopes
such as the $\it HST$. However (to our knowledge), no convincing CIV detection currently exists for targets whose sight-lines
are wholly contained within the confines of the LC. Although the apparent dearth of detectable CIV within the LC could be explained by the
existence of tangential magnetic fields emanating from the walls
of the LC that can quench the production sites of conduction \citep{cox03}, it is difficult to believe that this should apply to all
local clouds whose sight-lines have been sampled through many UV absorption measurements towards numerous
stars in the 50 - 150pc range. We note that $\it possible$ detections of CIV absorption originating within the LC
have been claimed for several targets located at distances beyond the boundary of the LC. These cases include
weak CIV absorption detected towards both $\beta$ CMa
(d= 153pc) and $\epsilon$ CMa (d = 132pc) at an absorption velocity consistent with that of the LIC \citep{dup98,gry01}.
However, several authors have questioned whether this highly ionized gas arises in a more distant cloud whose velocity
is similar to that of the LIC \citep{heb99,wood02}. In addition, CIV absorption was
detected at the velocity of the G-cloud towards two B-stars (d $\sim$ 180pc) in the direction of the Loop I bubble
by Welsh $\&$ Lallement \cite{wel05}, but the
association with very local absorption was discounted on the grounds that highly ionized gas was not detected towards two closer 
(d $<$ 130pc) stars with similar sight-lines. 
Furthermore, we note that although the local ISM has been well sampled
using hot white dwarf stars as background UV continuum sources, no detections of interstellar NV or CIV have been
claimed, although detections of both photospheric and circumstellar NV and CIV currently exist \citep{ban03}.

In this Letter we report on the detection of highly ionized interstellar CIV absorption that unambiguously arises in a gas cloud
located $\it within$ the Local Cavity along the sight-line towards the star HD 158427 ( d = 74pc). Based on the velocity
of one of the absorption components, the highly ionized absorption can be physically associated
with the Local Interstellar Cloud. This detection therefore helps constrain
the possible ionization mechanisms present within the Local Cavity.
 
\section{Observations and Data Reduction}
Observations of the B2Ve star, HD 158427 (d = 74$\pm$5pc) were made under the NASA $\it HST$ Cycle 17 
Cosmic Origins Spectrograph Guaranteed Time observation program GTO-11525 in October 2009, using the recently re-furbished STIS instrument \cite{wood98}. The
data were recorded through the 0.2 x 0.05 arcsec aperture (with neutral density filtering) using the E140H
echelle grating. Ultraviolet photons were collected with the far UV Multi-Anode Microchannel Array (MAMA) detector during two
exposures of 1760s  and 1838s with the grating centered at 1343\AA\ and 1598\AA\ respectively.
This resulted in spectra recorded at a resolving power of R $\sim$ 114,000 (2.6 km s$^{-1}$) for the
spectral orders observed over the entire 1245 - 1730\AA\ region. The data were processed at the Multi-mission Archive of the Space Telescope
Science Institute (MAST)
using the CALSTIS pipeline software with the final spectra being presented in the heliocentric scale with a typical accuracy of
$\sim$ $\pm$1 km s$^{-1}$.

Data for the CIV doublet ($\lambda$1548.22\AA\ and $\lambda$1550.77\AA) appear in two adjacent echelle orders,
which were co-added (and weighted inversely with respect to the continuum noise) to improve the resultant S/N of the data.
Local continua were established for both lines and their resultant intensity absorption profiles fit simultaneously with multiple
absorption components as described in Welsh $\&$ Lallement \cite{wel05}.  A best fit to the
 doublet was obtained using a 2-component cloud model
and in Figure 1 we show the observed residual intensity profiles
for both of the CIV absorption lines together with their component fit values of column density, N(CIV), doppler $\it b$-value and
cloud component heliocentric velocity, V.  

\section{Discussion}
\subsection{Is the CIV absorption of interstellar origin?}
Although we have clearly detected significant CIV absorption in the UV spectrum of HD 158427, we need to establish
that it is not of a stellar origin. Firstly, we can rule out a photospheric origin for CIV based on the extreme weakness of the
SiIII ($\lambda$1294\AA\  $\&$1299\AA) and CIII ($\lambda$1247\AA) photospheric lines in this spectrum.  Secondly, the CIV lines are
not formed at the known radial velocity of star (0 km s$^{-1}$), but instead one of the
two absorption components (i.e. V$_{1}$ $\sim$ -24.3 km s$^{-1}$) is
formed within $\pm$ 2 km s$^{-1}$ of the central absorption velocities of the low ionization
UV (SII, SiII, NiII and FeII) and optical (NaI and CaII)  interstellar lines
observed towards this star by Welsh $\&$ Lallement \cite{welsh10b}. We show the SII ($\lambda$ 1259\AA)
 and FeII ($\lambda$1608\AA) absorption
line profiles in Figure 1.
Thirdly, the high rotational velocity of the star (V$_{rot}$ $\sim$ 375 km s$^{-1}$) also precludes a stellar origin for the observed
narrow CIV absorption lines. Therefore, we believe that at least one of the observed CIV absorption components, V$_{1}$,  is interstellar.

\subsection{Does the CIV absorption arise within the Local Cavity?}
In Figure 2 we show the position of HD 158427 with respect to the spatial distribution of neutral gas (as traced by NaI absorption)
within 100pc of the Sun \citep{welsh10}. The star is clearly placed well within the dense surrounding neutral boundary of the LC. The
velocity vectors of  any absorption arising in the LIC and G-clouds \citep{lall95}, as projected towards HD 158427, are -23.1 km s$^{-1}$ and
-26.5 km s$^{-1}$ respectively. Thus, component V$_{1}$ can confidently be assigned to absorption arising within the LIC complex. Additionally,
the column density of this component, log N(CIV) = 12.20 cm$^{-2}$ is very similar to that assigned to LIC absorption
in the sight-lines towards both $\beta$ CMa
and $\epsilon$ CMa \citep{dup98,gry01}.
We also derive an upper limit of log N(SiIV) $\le$ 11.25 cm$^{-2}$ for the HD 158427 sight-line, such that the
observed ratio of N(CIV)/N(SiIV) $>$ 8.9 is consistent with the predictions
of cloud evaporation photoionization models \citep{slav02}. 

The V$_{2}$ = -41.3 km s$^{-1}$ component observed in both of the CIV lines is not seen in
any of the lower ionization UV species. Interstellar gas with such a
relatively high velocity is rare within the LC and its presence could be explained if it was associated with an outflow from the nearby
Sco-Cen/ Loop I superbubble ($\it l$ = 330$^{\circ}$, $\it b$ = +18$^{\circ}$).  This X-ray emitting interstellar cavity (d $\sim$ 130pc)
is thought to be produced by the collective stellar
winds of the Sco-Cen OB association and several consecutive SN events, such that this bubble is expanding towards the LC with (hot)
gas presumably flowing into the LC through the fragmented gap in the boundary wall shown in Figure 2. Support for this view comes from
soft X-ray images that show an interaction region between the two interstellar cavities \citep{egg98} and the fact that highly ionized gas with 
a velocity of -34 km s$^{-1}$ has been detected flowing away from Loop I by Welsh $\&$ Lallement \cite{wel05}.

\subsection{Ramifications of the CIV detection}
High ions produced at an evaporating conductive interface between hot and cooler interstellar gas should possess doppler line-widths
that correspond to a thermal gas temperature of $\sim$ 10$^{5}$K (i.e. $\it b$ $>$ 10 km s$^{-1}$). We derive a gas temperature of
$\sim$65,000K ($\it b$ = 9.5 km s$^{-1}$) for the V$_{1}$ component, which is consistent with what one might expect for the CIV ion.
However, the doppler line width of the  V$_{2}$  component is indicative of an unusually low gas temperature of $\le$ 34,000K. 
Narrow doppler-widths
can occur for highly ionized gas that is cooling rapidly and out of equilibrium or if it is associated with weak shocks. 
Similarly low gas temperatures have been obtained for the SiIV and CIV lines seen by Welsh $\&$ Lallement \cite{wel05} towards the Loop I superbubble.

Although we have detected highly ionized gas associated with the LIC, we note that many other UV observations towards nearby stars have
failed to do so, which in many cases may well be due to instrumental sensitivity limitations. 
The local distribution of the OVI ion has
been described as `patchy' \citep{sav06}, and this may also apply to the spatial distribution of local CIV (and NV) ions.
Although detailing
specific theoretical scenarios that might explain this observed `patchiness' is beyond the present scope of this paper, we offer some speculative
thoughts on this point. For example, if CIV is generated within turbulent mixing layers (which will depend on an existing shear flow), then
a non-unifrom flow pattern in the surrounding hot gas could pull off some cold/warm material from the LIC which then mixes and cools and
produces localized CIV gas. Alternately, the geometry of the LIC could be such that it only receives ionizing (soft X-ray) photons along specific
sight-lines, dependent on the 3-D distribution of the intervening opacity of the local ISM. Finally, one can always invoke a
peculiar magnetic field
topology to control the gas evaporation through thermal conduction.

\section{Conclusion}
We have presented high resolution UV absorption measurements of the highly ionized CIV doublet at $\lambda$1550\AA\ using the
refurbished $\it HST$ STIS instrument.  Both line-profiles require a two-component cloud structure to fit these data, with one
cloud component (V$_{1}$) arising at the projected velocity of LIC (-24.3 km s$^{-1}$) and the other (V$_{2}$) at -41.3 km s$^{-1}$. Component 
V$_{1}$ can almost certainly be associated with absorption arising within a layer of 65,000K gas located within 5pc of the Sun (in the LIC) and its observed CIV column density agrees well with theoretical predictions for the production of this ion in the LIC. However, the doppler width required
to fit the V$_{2}$ gas cloud, which we believe may be associated with an outflow of
hot and highly ionized gas from the nearby Loop I superbubble,  suggests a low temperature of $\sim$ 34,000K for the CIV gas.
These data are part of far larger program of UV absorption observations of local interstellar gas currently being carried out under the
Cosmic Origins Spectrograph Science team's guaranteed observer programs using both the $\it HST$-STIS and
$\it HST$-COS instruments during $\it HST$ Cycles 17 and 18. Therefore, as we receive more
of thee observations we believe that they
may provide answers to some of the oustanding problems raised in this paper, particularly with respect to the apparent
`patchy' spatial distribution of highly ionized gas absorption within 100pc.

\begin{acknowledgements}
We particularly acknowledge all of the dedicated
team of engineers, technicians, scientists and astronauts who contributed to the success of the
the STS-125 servicing mission to the Hubble Space Telescope. We thank Jonathan Slavin and John Vallerga for very
useful discussions, and also thank the $\it HST$-COS science team for their advice and financial support through
NASA GSFC grant 005118.

\end{acknowledgements}

\begin{figure*}
\center
\plotone{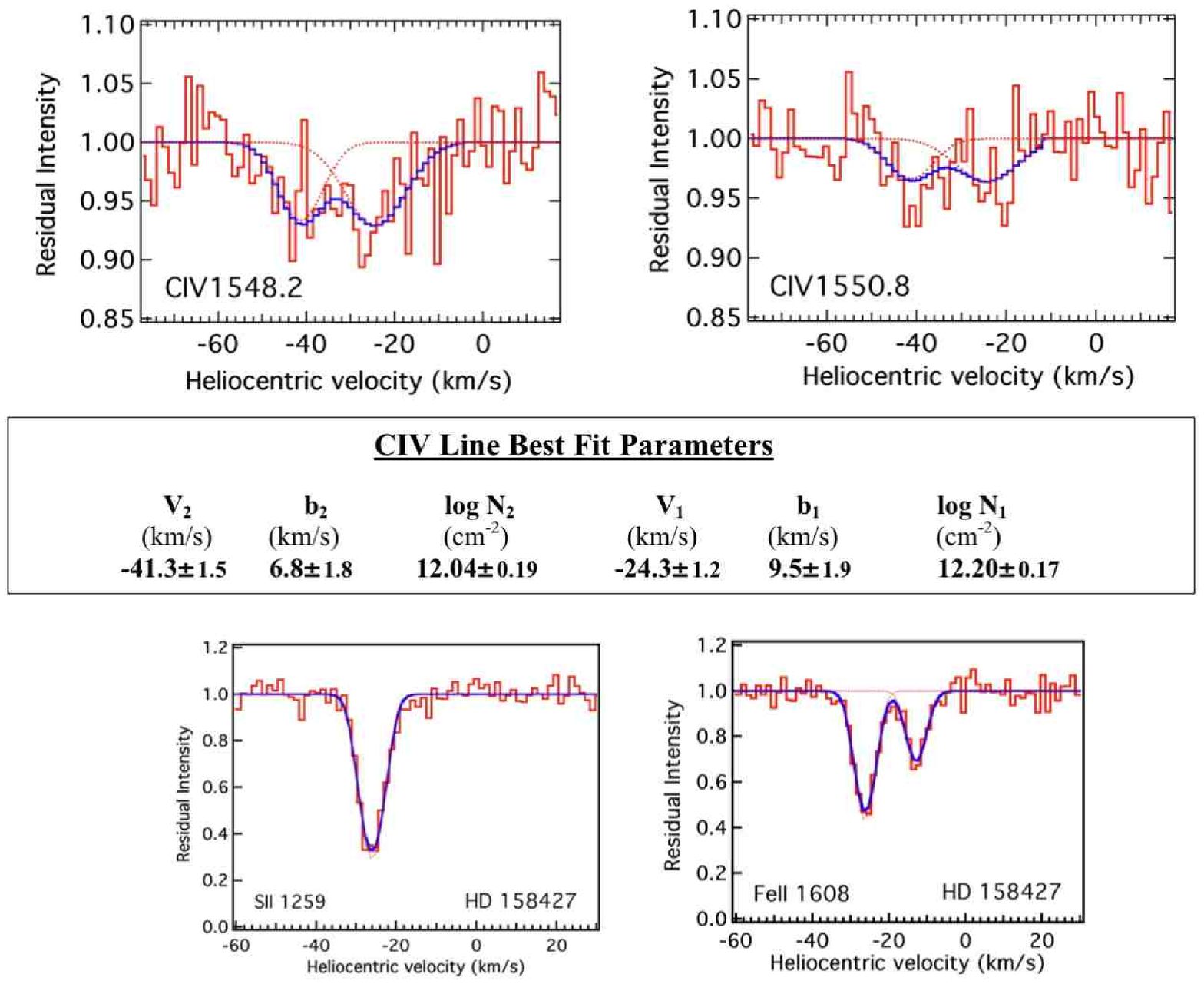}
\caption{Interstellar absorption profiles of the CIV ($\lambda$1548 and $\lambda$1550\AA) lines observed towards HD 158427. Full lines are the model fits superposed upon the normalized residual intensity data points. Dotted lines are the unconvolved components. The best-fit parameters for the two-cloud component fit for both CIV lines (which were fit simultaneously) are listed in the box. Also shown are the low ionization interstellar lines of SII1259\AA\ and FeII1608\AA\ recorded towards this star. }
\label{Figure 1}
\end{figure*}

\begin{figure*}
\center
\epsscale{0.5}
\plotone{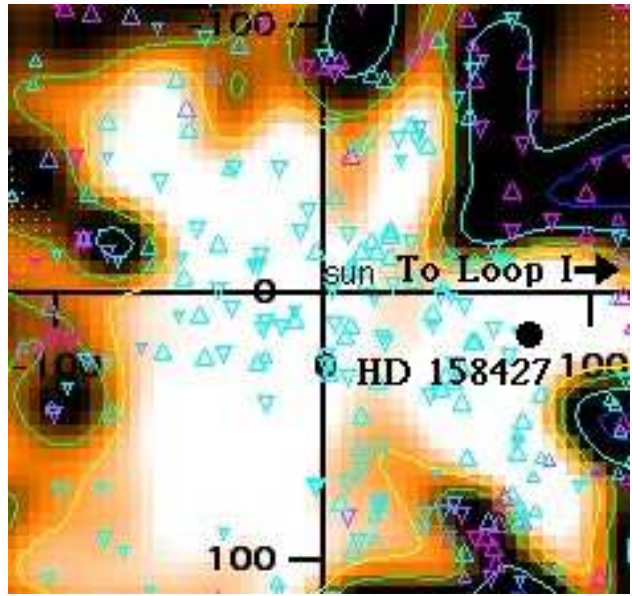}
\caption{The position of the HD 158427 sight-line with respect to the 3-D spatial contours of the neutral NaI gas within $\pm$100pc of the Sun
as viewed in the galactic plane (Welsh et al. 2010). Dark shading represents dense and cold absorbing gas, whereas light shading represents regions of minimal neutral gas density. Small triangles represent stellar targets whose absorption data were used to construct the map. }
\label{Figure 2}
\end{figure*}

\end{document}